# Visual Infrared Video Fusion for Night Vision using Background Estimation

Anjali Malviya, S. G. Bhirud

**Abstract**— Video fusion is a process that combines visual data from different sensors to obtain a single composite video preserving the information of the sources. The availability of a system, enhancing human ability to perceive the observed scenario, is crucial to improve the performance of a surveillance system. The infrared (IR) camera captures thermal image of object in night-time environment, when only limited visual information can be captured by RGB camera. The fusion of data recorded by an IR sensor and a visible RGB camera can produce information otherwise not obtainable by viewing the sensor outputs separately. In this paper we consider the problem of fusing two video streams acquired by an RGB camera and an IR sensor. The pedestrians, distinctly captured by IR video, are separated and fused with the RGB video. The algorithms implemented involve estimation of the background, followed by detection of object from the IR Video, after necessary denoising. Finally a suitable fusion algorithm is employed to combine the extracted pedestrians with the visual output. The obtained results clearly demonstrate the effectiveness of the proposed video fusion scheme, for night vision.

—————————— ◆ ——————————

## 1 INTRODUCTION

THE term fusion in general means an approach to extract information acquired in several domains. Video Fusion is the process of combining relevant information from two or more videos into a single video. The resulting video will be more informative than any of the input video [1]. The goal of video fusion is to integrate complementary multi-sensor, multi-temporal and/or multi-view information into one new video containing information, the quality of which cannot be achieved otherwise. We perform multi-sensor fusion of visual and infrared (IR) videos. In night-time environment, only limited visual information can be captured by CCD cameras under poor lightning conditions, thus making it difficult to do surveillance only by visual sensor. Meanwhile IR camera, that is IR sensor, captures thermal image of object. Thermal image of pedestrian in night-time environment can be seen clearly in infrared video sequence used for this work. Infrared video provides rich information for higher temperature objects, but poor information for lower temperature objects. Visual video, on the other hand, provides the visual context to the objects [2]. Thus, the fusion of the two videos will provide good perceptibility to human vision under poor lightning condition. This will help detect the moving objects (pedestrian in our video) during night-time.

————————————————

- *Anjali Malviya is working as Asst. Professor with the Dept. of IT, TSEC, University of Mumbai, India.*
- *S. G. Bhirud. is working as Asstt. Professor with the Dept. of Computer, VJTI, Mumbai, Indi.a..*

## 2 INFRARED IMAGING

In infrared photography, the film or image sensor used is sensitive to infrared light. The part of the spectrum used is referred to as near-infrared to distinguish it from far-infrared, which is the domain of thermal imaging. Wavelengths used for photography range from about 700 nm to about 900 nm. Usually an infrared filter is used; this lets infrared light pass through to the camera, but blocks all or most of the visible light spectrum.

When these filters are used together with infrared-sensitive film or sensors, very interesting in-camera effects can be obtained; false-color or black-and-white images with a dreamlike or sometimes lurid appearance. Infrared light lies between the visible and microwave portions of the electromagnetic spectrum. Infrared light has a range of wavelengths, just like visible light has wavelengths that range from red light to violet. "Near infrared" light is closest in wavelength to visible light and "far infrared" is closer to the microwave region of the electromagnetic spectrum. The longer, far infrared wavelengths are about the size of a pin head and the shorter, near infrared ones are the size of cells, or are microscopic.

### 2.1 Thermal Imaging

Infrared thermography or thermal imaging or thermal video, is a type of infrared imaging science. Thermographic cameras detect radiation in the infrared range of the electromagnetic spectrum (roughly 900–14,000 nanometers or 0.9–14 μm) and produce images of that radia-tion, called thermograms [Fig. 1]. Since infrared radiation is emitted by all objects based on their temperatures, according to the black body radiation law, thermography makes it possible to see one's environment with or without visible illumination. The amount of radiation emitted by an object increases with temperature, therefore thermography allows one to see variations in temperature and hence the name. When viewed by thermographic camera, warm objects stand out well against cooler backgrounds; humans and other warm-blooded animals become easily visible against the environment, day or night [Fig. 2]. Some physiological activities, particularly responses, in human beings and other warm-blooded animals can also be monitored with thermographic imaging.

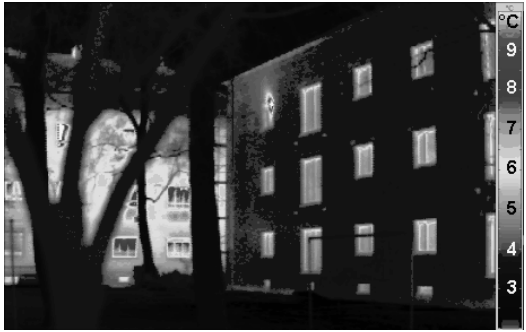

Fig. 1.Thermogram of buildings, with an approximate Celsius scale

## 3 MOVING OBJECT DETECTION

During night time, under poor lighting conditions, the CCD camera captures only limited visual information, thus making it difficult to do automatic video surveillance only by CCD camera, meanwhile, an infrared camera captures thermal image of an object, that is, it provides high amount of information for higher temperature objects but few information for lower temperature objects. For example, in a visual video, a car in the brighter region is seen clearly but a pedestrian in the darker region can hardly be found, while on the other hand, in infrared video, the pedestrian with higher temperature than the surroundings is clearly seen but the sidewalk and buildings can hardly be found [3]. Using two types of sensors therefore will be ideal for enhanced visibility during night time. Noise in IR video is a severe problem. If neglected, it will have undesirable effect on the surveillance application. The sources of such noise may vary from thermal emitters in the environment to temperature variation around IR sensor. Figure 3 shows the flowchart of the proposed video system.

### 3.1 Background Estimation
In this work, pre-aligned thermal and visual videos are taken [4]. The popular median filtering algorithm, which sets the estimated background equal to the median value of the input frames, assumes that the background is visible for at least half the filter length. In [5] the extraction process computes a map analyzing the history of each pixel, in terms of temporal stability and variability. Background estimation is performed on the input sequences. We take the frames at every T/4 seconds and pass each frame through a median filter. A table is maintained in the form of array, with rows columns and frame number, giving the value with each combination. The maximum occurring pixel value is considered for each pixel and put it in the new image which becomes our background.

### 3.2 Object Extraction
The moving object is extracted by subtracting the estimated background from the IR frames. We subtract the estimated background image from each IR frame and if the pixel value is greater than predeciced threshold then we consider it as 255 else 0. Thus we obtain a binary mask. The brightness-only information is not robust in detection of moving object, as there are other thermal emitters in the environment which can lead to false detection. In order to improve the robustness of detection, we also take into consideration, area and shape information [7]. If the area and *height/width* ratio of the detected object is in the acceptable range, then the region is regarded as a human else it is regarded as noise region by environmental thermal emitting [9]. For each detectede region in the image, using the bounding box property, we obtain the height, width and the x and y co-ordinates of the top left corner of the minimum area rectangle that completely encloses the candidate region. Set the values of area threshold, minimum height/width ratio, and maximum height/width ratio. For each of the identified candidate region we perform a check on the threshold and classify the detected region as object region.

### 3.3 Fusion of Object and Background
At the end of the detection method, infrared video frames are fused with visual video frames to provide visual context [8][10]. The first stage involves extraction of the RGB components of pixels in visual video, of the locations where the object was identified in IR video. We then add a suitable threshold to the RGB components.

## 4 RESULTS

The input dataset comprised of the visible and IR videos [source: http://www.imagefusion.org/]. This dataset consists of two video sequences, one in the visible spectrum and the other in thermal infrared. Both are compressed into AVI format and contain 527 frames each. The frame rate of each video is 23.96fps. The videos were decomposed into IR frames [Fig.3] and visible frames [Fig. 4]. The result obtained at every stage is shown via four frames of the video streams. The segmentation of the objects from the IR video is followed by denoising. Finally the detected moving objects are fused with the visible camera frames. A bounding box has been displayed around the pixels obtained in the object regions to clearly demarcate it in the fused frames [Fig. 5].

## 5 CONCLUSION

An obvious requirement for surveillance system is real time performance. If the system has to process signals from multiple sensor and modalities, then the required processing is multiplied. Moreover requirement for robustness and accuracy tend to make the algorithm design complex and highly computational.

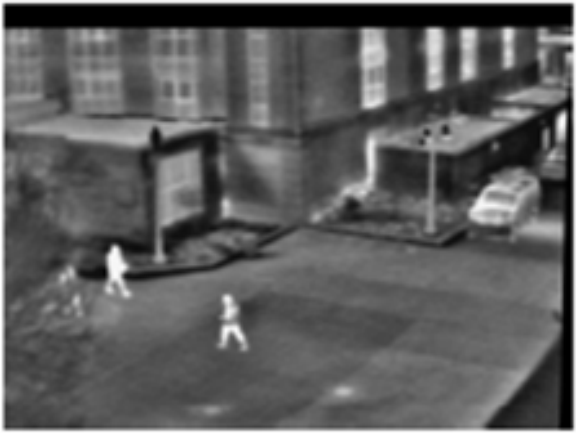

Fig. 2. Thermal image of the scene

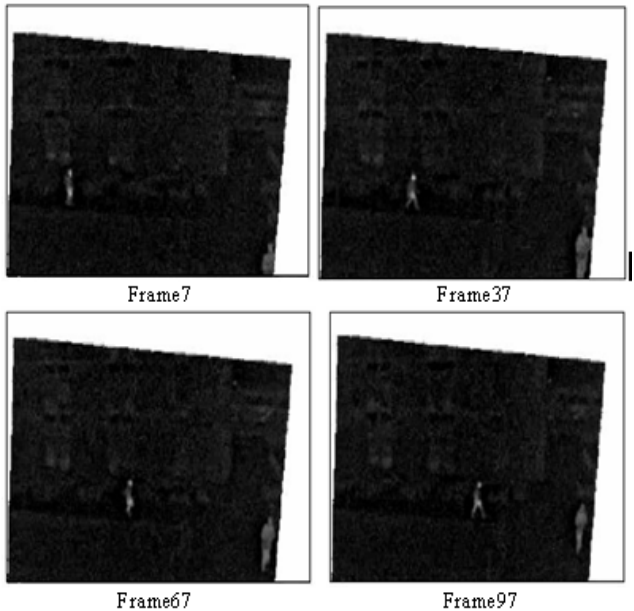

Fig.3. Frames from the IR video

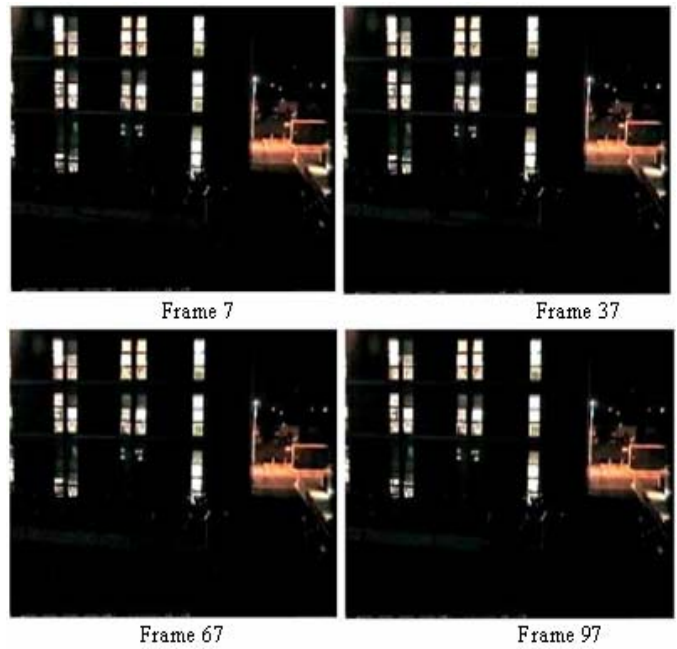

Fig. 4. Frames from the Visible video

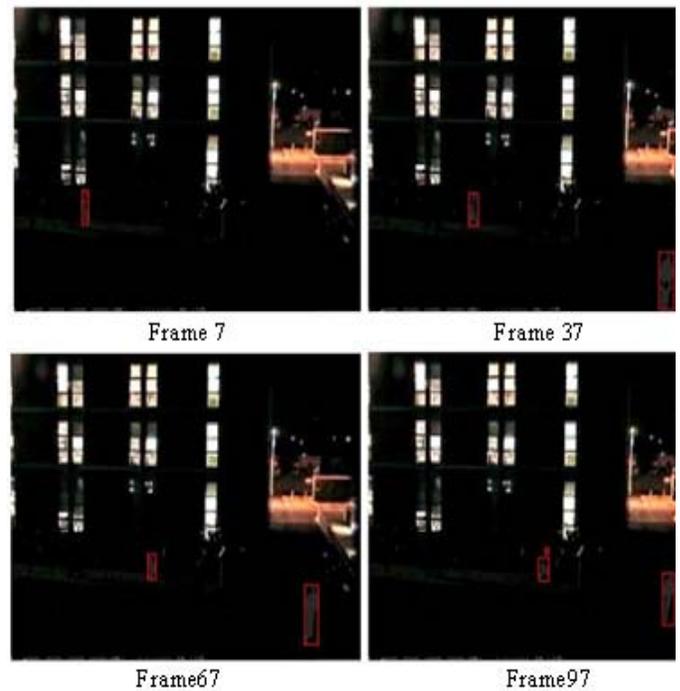

Fig. 5. Objects from IR frames, fused on Visible Camera frames

## REFERENCES

[1] C. Pohl, J. L. Genderen, "Multisensor Image Fusion in Remote Sensing: Concepts, Methods and Applications", International Journal of Remote Sensing, vol. 19, no. 5, pp. 823-854, 1998

[2] Y.Chen, C.Han, "Night-time Pedestrian Detection by Visual-Infrared Video Fusion." Proceedings of 7th World congress on Intelligent Control and Automation, June 25-27, 2008, Chongoing, China.

[3] J.Davis, V.Sharma, "Fusion-Based Background-Subtraction using Contour Saliency." CVPR Workshop OTCBVS, 2005.

[4] P.A.Viola. "Alignment by Maximization of Mutual Information." Phd thesis, Massachusetts Institute of Technology, Massachusetts (MA), USA, June 1995

[5] J. Zhang, L. Zhang, H. Tai, "Efficient Video Object Segmentation Using Adaptive Background Registration and Edge-Based Change Detection Techniques", IEEE International Conference on Multimedia an Expo (ICME), pp. 1467-1470, 2004.

[6] R. Adams and L. Bischof, "Seeded region growing," IEEE Trans. Pattern Anal. Machine Intell., vol. 16, no. 6, pp.641C647, 1994.

[7] J.Foster, M. Nixon, and A.Prugel-Bennett, "Automatic Gait Recognition Using Area-Based Metrics, Pattern Recognition," Letters, Vol. 24, pp.2489-2497, 2003

[8] A. Toet, L. J. Van Ruyven, J. M. Valeton, "Merging Thermal and Visual Images by Contrast Pyramid", Optical Engineering, vol. 28, no. 7, pp. 789-792, 1989

[9] Toet, "Multiscale Contrast Enhancement with Application to Image Fusion", Optical Engineering, vol. 31, no. 5, pp. 1026-1031, 1992

[10] S. Jabri, Z. Duric, H. Wechsler, A. Rosenfeld, "Detection and Location of People in Video Images Using Adaptive Fusion of Color and Edge Information", 15th International Conference on Pattern Recognition (ICPR'00), vol. 4, pp. 4627- 4630, 2000)